# Icosahedral packing of polymer-tethered nanospheres and stabilization of the gyroid phase

Christopher R. Iacovella[1], Aaron S. Keys[1],
Mark A. Horsch[1], and Sharon C. Glotzer[1,2,*]

[1]Department of Chemical Engineering and
[2]Department of Materials Science & Engineering

8 February 2007

We present results of molecular simulations that predict the phases formed by the self-assembly of model nanospheres functionalized with a single polymer "tether", including double gyroid, perforated lamella and crystalline bilayer phases. We show that microphase separation of the immiscible tethers and nanospheres causes confinement of the nanoparticles, which promotes local icosahedral packing that stabilizes the gyroid and perforated lamella phases. We present a new metric for determining the local arrangement of particles based on spherical harmonic "fingerprints", which we use to quantify the extent of icosahedral ordering.

The ability of block copolymers (BCPs) to order into periodic micro-domains has made them a choice building block for a wide variety of applications ranging from drug delivery [1] to photonic-bandgap materials [2, 3]. In particular, the BCP bi-continuous phases, when appropriately modified with nanoparticles, are seen as ideal candidates for catalytic materials and high conductivity nanocomposites [4]. The use of polymer-tethered nanoparticles is a novel strategy for the self-assembly of ordered arrays of nanoparticles where the bulk phases formed resemble the complex morphologies found in BCPs and surfactants [5, 6]. Polymer-tethered nanoparticles constitute a class of "shape-amphiphiles" where microphase separation occurs due to the immiscibility between the nanoparticle and tether [7]. In these systems, the geometry of the nanoparticle heavily influences the bulk structure and local arrangement of nanoparticles by inducing liquid crystalline ordering [5, 7]. Previous simulation work on polymer-tethered nanorods highlights the importance of the interplay between microphase separation and particle geometry; e.g., the phase behavior for tethered rods includes a chiral cylinder phase and smectic lamellar phases with ordered perforations [5], which are not observed in flexible BCPs. Additional work in the literature suggests that morphologies may adopt unique structures as a result of confinement, including helical structures formed by colloids confined in v-shaped grooves [8] and helices and tori formed from BCPs confined in cylindrical pores [9].

In this work, we examine the bulk-phase microstructures formed by polymer-tethered nanospheres (TNS) with attractive particles and repulsive tethers and demonstrate via simulation the first nanoparticle-based double gyroid phase. We present a new metric of local order based on spherical harmonics and use this to explore the impact of microphase-separation induced confinement on the local ordering of spherical particles. We show that this confinement promotes icosahedral packing of hard attractive particles, which helps to stabilize certain microphase-separated structures with limited stability in BCP systems, including the double gyroid.

---

[*] Corresponding author: sglotzer@umich.edu

To allow for the realization of long time scales and large systems required to self-assemble complex mesophases from initially disordered systems, we use the method of Brownian dynamics (BD). In BD, each bead is subjected to conservative, frictional and random forces $\mathbf{F}_i^C$, $\mathbf{F}_i^F$, and $\mathbf{F}_i^R$, respectively, and its trajectory obeys the Langevin equation [10], $m\ddot{\mathbf{r}}_i = \mathbf{F}_i^C + \mathbf{F}_i^F + \mathbf{F}_i^R$. To consider a general class of tethered nanoparticles rather than any one specific system, we implement empirical pair potentials that have been successful in the study of BCPs and surfactants. Nanospheres are modeled as beads of diameter $2.0\sigma$ permanently connected to tethers via finitely extensible non-linear elastic (FENE) springs [11]. Tethers are modeled as bead-spring chains containing eight beads of diameter $\sigma$ connected via FENE springs. To model the attractive interaction between nanoparticles, we use the Lennard-Jones potential (LJ), where particle-particle interactions are shifted to the surface. Solvophilic tethers and species of different type interact via a purely repulsive Weeks-Chandler-Anderson soft-sphere potential to account for short-range, excluded volume interactions. A schematic of the building block can be found in figure 1. The degree of immiscibility and solvent quality are determined by the reciprocal temperature, $\varepsilon/k_BT$, where $\varepsilon$ is the LJ well depth. The natural units of these systems are $\sigma$ and $\varepsilon$ and the dimensionless time is $t^* = \sigma\sqrt{m/\varepsilon}$, where $m$ is the mass of a tether bead, $\sigma$ is the diameter of a tether bead, and $\varepsilon$ is the interaction parameter. In all cases, volume fraction, $\phi$, is defined as the ratio of excluded volume of the beads to the system volume, and dimensionless temperature, $T^*$, is defined as $k_BT/\varepsilon$. Further details of the model and method can be found in references [6, 7]. This model relates well to experimentally synthesized building blocks including polymer-functionalized fullerenes [12, 13], dendrimer functionalized fullerenes [14], tethered nanoparticles formed by crosslinking one block of a BCP [15], and tethered quantum dots [16].

*Bulk ordering:*

To examine the bulk phase ordering of the TNS system with attractive nanoparticle headgroups, we explore the temperature vs. volume fraction phase diagram for a system composed of TNS building blocks with eight-bead polymer tethers affixed to nanoparticles of diameter of $2\sigma$. Simulations are conducted at ten fixed volume fractions between 0.15 – 0.45. Systems are prepared in a high temperature disordered state and incrementally cooled until bulk ordered phases are reached, as determined by visual inspection and discontinuities in potential energy as a function of temperature [17]. For each volume fraction, multiple runs are conducted at various cooling rates and system sizes to avoid dynamically trapped structures and to mitigate finite size effects. The results presented here are based on ~40 independent simulation runs of ~250 state points each at values of $T^*$ between 0.21 to 2.0, for systems ranging from 500 to 4000 tethered nanoparticles, or 4500 to 36000 individual beads, respectively.

A phase diagram summarizing the observed phases is presented in Fig. 1. Each data point in the figure represents a state point arrived at using multiple cooling rates and system sizes as described above. With increasing volume fraction, we observe disordered wormy micelles (DWM), hexagonally packed cylindrical micelles (H), the bi-continuous double gyroid (DG), perforated lamellae where the perforations are through the nanosphere layer (PLH), and lamellar bilayers (L). At temperatures above the order-disorder temperature, $T_{ODT}$, we find disordered aggregates rather than periodic bulk structures and at sufficiently high temperatures we find no aggregates.

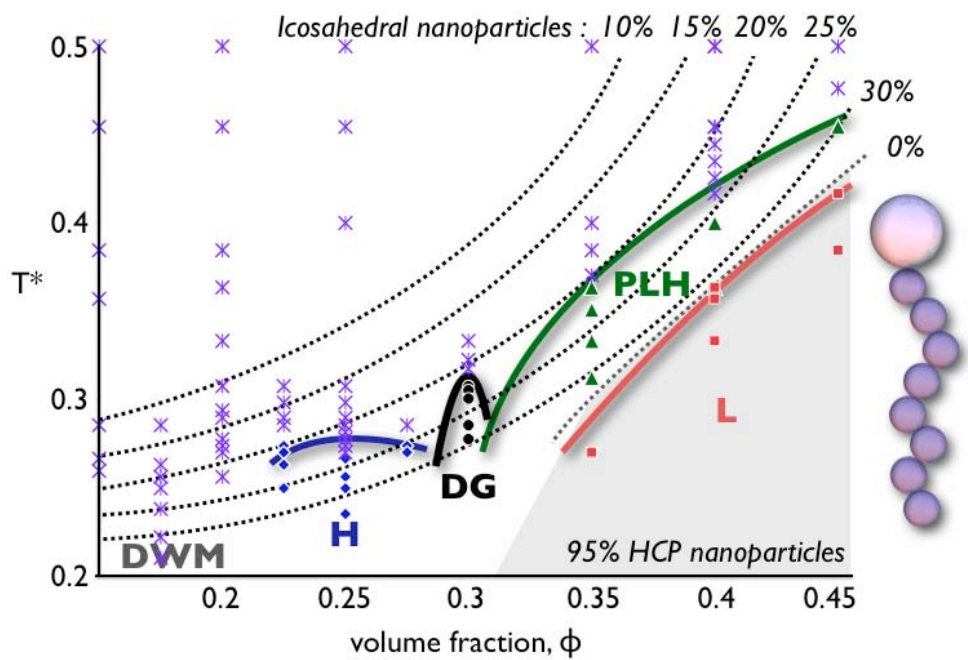

*Figure 1:* Temperature vs. volume fraction phase behavior, where solid lines represent approximate phase boundaries determined by ~250 state points spanning $\phi$ =0.15-0.45 and $T^*$=0.21-2.0; in almost all cases, simulations were conducted using multiple system sizes to mitigate finite size effects. Stars indicate simulated points in the disordered region of the phase diagram. With increasing volume fraction we observe disordered wormy micelles (DWM), hexagonally packed cylinders (H), double gyroid (DG), perforated lamellae where perforations are through the nanoparticles (PLH), and lamellar bilayers (L). Dotted lines fit data points indicating values of T and $\phi$ at which the indicated percentages of icosahedral clusters formed by nanoparticle headgroups are found. The shaded region indicates the range of $T^*$ and $\phi$ over which crystalline ordering of the nanoparticles is observed. A schematic of the model tethered nanosphere is shown at right.

Interestingly, we observe both DG and PLH phases between $\phi$ ~0.30 – 0.45. A simulation snapshot of the DG phase is shown in Figure 2a and a simulation snapshot of the PLH phase is shown in Figure 2b. In BCP systems, both the DG and PLH phases have been reported in similar regions of the phase diagram [18]; however, the DG phase is considered to be an equilibrium morphology [19] and the PLH phase is considered to be metastable, stabilized by compositional fluctuations [20]. The DG phase has been widely found using lattice based simulation methods [21, 22], but few examples of this complex phase exist for the simulation of BCPs and surfactants using molecular dynamics based methods [23, 24] making this result unusual.

For the TNS system, the dominance of the PLH phase outside of the DG region suggests that over this concentration range it is stable and reproducible; we did not observe bi-continuous structures within the PLH region. As in the case of the polymer-tethered rods, the perforations are a result of a competition between the tendency for the nanoparticle to locally order and the tendency for the tether to maximize its configurational entropy [5].

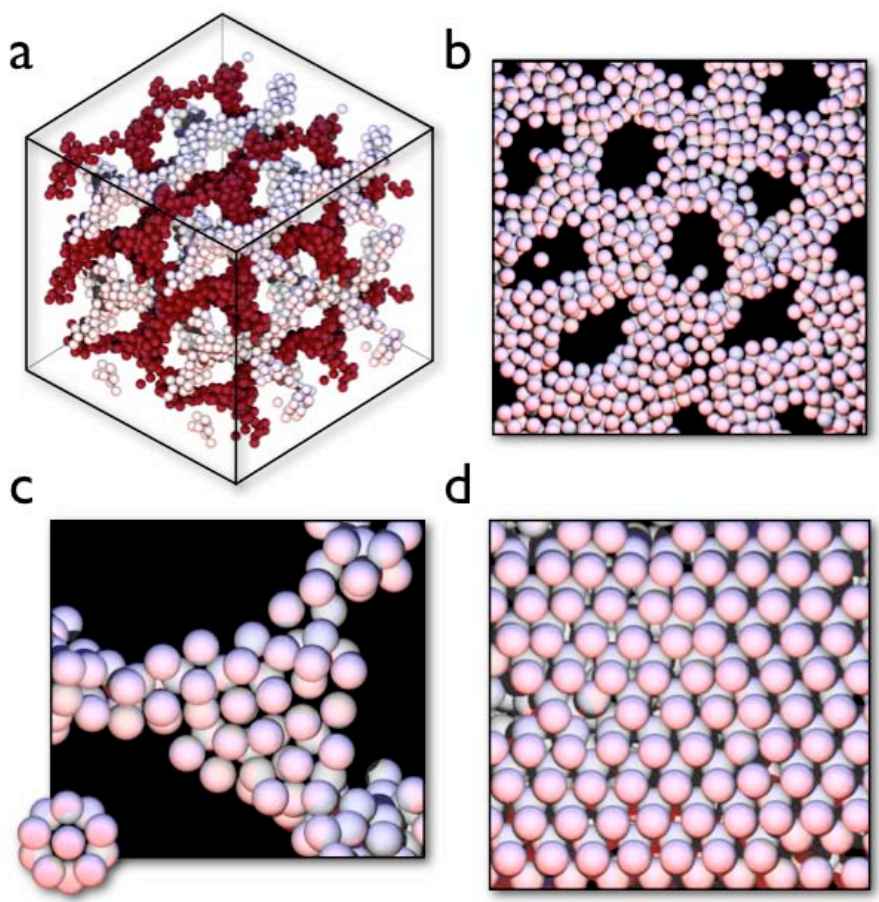

*Figure 2: Images of self-assembled structures.* In all cases, tethers have been removed for clarity. (a) Double gyroid phase; the minimal unit cell was duplicated and found to be stable over ~10 million time steps. (b) Individual sheet of perforated lamella, showing a tendency for the perforations to adopt a honeycomb packing. (c) Node of the gyroid, showing icosahedral rings; perfect icosahedron inset. (d) Crystalline packing of lamellar bilayer.

**Stability of the gyroid:** The limited stability of the DG phase in BCP systems has been attributed to packing frustration within the connection points (nodes) [25, 26], which arises due to a high void fraction (low packing density) within the nodes. To examine the void fraction within the nodes, we look at relative trends in void fraction. We approximate the center of a node and calculate the void fraction within a spherical volume drawn from the center, repeating for various sphere radii, $r_{cut}$ where $r_{cut}$ is in units of $\sigma$. We find that for small values of $r_{cut}$ the void fraction is lower than in the bulk, starting at ~0.44, and as $r_{cut}$ is increased we approach the bulk void fraction of ~0.7 as shown in Figure 3. This shows that particles within the nodes are packed more densely than in the bulk. Martinez-Veracoechea and Escobedo [16, 17] used a similar analysis to compare a monodisperse BCP system to a blend of two different length BCPs. In the blend system, the authors found that the longer of the two polymers occupied the nodes of the double gyroid, resulting in a larger range of stability compared to the monodisperse system [22]. For the monodisperse system, the authors found that the void fraction is higher than the bulk for small values of $r_{cut}$, and decays to the bulk value as $r_{cut}$ becomes large [22]. In the case of the blend, the authors found that, like our system, the void fraction is *lower* than the bulk for small values of $r_{cut}$, and approaches the bulk value from *below* as $r_{cut}$ becomes large [22]. This similarity in trends suggests that the ability of the TNS system to realize the DG phase is linked to the ability of nanoparticles to locally order into dense structures, decreasing the packing frustration.

By visual inspection, we find evidence that nanoparticles form ring-like structures resembling icosahedral clusters at the nodes, as shown in Figure 2c. Unlike icosahedral ordering observed in dense liquids, the tether sterically restricts the way in which the particles can pack, resulting in clusters with only partial coordination.

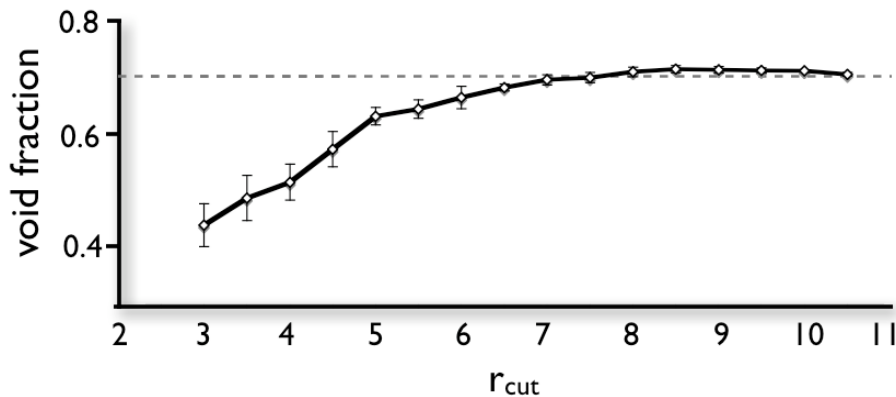

*Figure 3:* Void fraction of a node vs. $r_{cut}$. For small $r_{cut}$ values, void fraction is lower than the bulk, approaching the bulk value of 0.7 as $r_{cut}$ is increased. The bulk void fraction of 0.7 is represented by the dotted line.

*New method for characterizing local structures:*

To characterize rigorously the local particle packing, we introduce a residual minimization scheme as a modification of the cluster shape-matching algorithm of Steinhardt, et al. [27]. For a thorough explanation of the scheme of Steinhard et al., see references [27-29]. Our scheme is general and can be applied to any liquid or crystalline system of spherical particles obtained via experimental or computational methods.

Our overall goal is to determine if a set of vectors *r* drawn from a particle to its nearest-neighbors matches a given reference structure (e.g. an icosahedron or a small face-centered cubic crystal cluster). Because non-periodic systems contain configurations with many different orientations, this analysis cannot be accomplished trivially by, e.g., applying a set of Cartesian vector dot products. The advantage of the spherical harmonics construction of Steinhardt, et al. is that it is capable of matching structures in a way that is independent of orientation.

To quantify the shape of a cluster two metrics of local ordering are defined: $Q_l$ and $w_l$. Mathematically, $Q_l$ is derived by evaluating a set of spherical harmonic functions $Y_{lm}(\theta, \phi)$ for each nearest-neighbor direction $\theta_i(r_i)$, $\phi_i(r_i)$. Here, *l* is the harmonic frequency and *m* is an integer that runs from –*l* to *l*. Summing over all bonds in the sample gives the *2l+1* dimensional vector $Q_l$, which contains one component for each *m*. Because of the symmetries of spherical harmonics, the magnitude of the $Q_l$ vector is invariant for a given cluster, regardless of its orientation. The quantity $w_l$ is derived by considering specific rotationally invariant combinations of the average values of the $Q_l$ vector components $Q_{lm}$.

The typical implementation of the Steinhard et al. scheme involves choosing one or two harmonics (usually *l=4* and *l=6*) and comparing the resulting $Q_l$ and $w_l$ values for the local structure to the expected values for one or more reference structures. In order to make this comparison it is necessary to define cutoff values that define a match (e.g., if $Q_6 > 0.6$ and $w_6 < -0.1$ then the shape matches an icosahedron [27]). This scheme works reasonably well for differentiating between two local structures [29], however, when attempting to differentiate between a large number of local structures this method becomes cumbersome, since cutoffs must be determined via trial and error. An additional subtle problem is that certain structures have similar values for particular harmonics (e.g. body centered cubic and simple cubic have the same $w_6$ value [30]). Thus, a given configuration may be misidentified when using such loosely defined cutoffs. This problem can sometimes be circumvented by using multiple harmonic frequencies *l*; however, this often requires that cutoffs be relaxed, again potentially resulting in misidentified structures.

To avoid these problems, we introduce a residual-based scheme, $R_{YLM}$, that does not rely on multiple cutoffs or substantial trial and error. The first step in our scheme is to define a set of reference structures with which we wish to compare and determine their $Q_l$ and $w_l$ values for *l* = 4, 6, … 12. We only consider even number harmonics because they are invariant under inversion and because these frequencies are the leading terms in the expansion [27]. To determine the local configuration of a given particle in our system, we calculate the residual value, $\chi_i = \sqrt{\sum_{l=4}^{12}(Q_l - Q_{ref_i})^2 + \sum_{l=4}^{12}(w_l - w_{ref_i})^2}$, with respect to each reference structure. A particle is considered to be in the local configuration *i* that minimizes the residual $\chi_i$. Of course, by construction, it is impossible to generate a reference structure for disordered local configurations. Thus, we classify a particle as belonging to a disordered local configuration if its residual exceeds a particular cutoff value, here chosen as 0.1. An advantage of this method is that only a single cutoff is required irrespective of the number of reference structures.

We find that for the monatomic 12-6 LJ system, the $R_{YLM}$ method is able to identify particles in a crystalline configuration with accuracy comparable to the local order parameter, $q_6 \cdot q_6$ [30], which has become a standard tool for analyzing crystals. An additional strength of the $R_{YLM}$ method is that it allows us to classify structures by more than just "crystal-" or "liquid-like" qualifiers; we can accurately determine the local arrangement of particles. The $R_{YLM}$ method is generally applicable to a wide range of systems, including, e.g., colloids and biopolymers as well as nanoparticles. We have validated the accuracy of this method for several model liquids; additional details of this method and its use will be presented elsewhere.

*Quantifying local ordering:*

To verify our visual findings, we incorporated into the $R_{YLM}$ reference database icosahedral clusters that maintain the same bond angles, but possess only partial coordination numbers (e.g. we remove 1 to 4 particle(s) from a perfect 13-particle icosahedral cluster). These clusters are indistinguishable from the minimum potential energy clusters found by Doye and Wales [31, 32]. Our reference library additionally contains other polyhedra and crystalline arrangements with full and partial coordination.

Applying the $R_{YLM}$ algorithm to the DG phase, we confirm that the local arrangements of nanoparticles are icosahedral with partial coordination as observed visually in Figure 2c. Applying this analysis to the entire temperature vs. volume fraction phase diagram, we observe that as temperature is decreased, there is an increase in the number of icosahedrally ordered nanoparticles with partial coordination corresponding to increased aggregation of nanospheres. Figure 1 shows the general trend of icosahedral ordering overlaid on the bulk phases; dotted lines indicate the set of values of temperature and volume fraction at which 10, 15, 20, 25, and 30% of the nanoparticles are central particles in icosahedral clusters. Each line is interpolated from the analysis of the available data points (points at 6 to 16 different temperatures for each volume fraction), where each point is averaged over ~10 different samples or ~10000 nanoparticles. While we see a strong increase in icosahedral ordering, we see very little increase in crystalline ordering until we reach the $T_{ODT}$ of the lamellar bilayer phase, at which point nanoparticles crystallize and the number of icosahedrally-ordered nanoparticles drops to nearly 0% (shaded area in Figure 1). This region of low icosahedral ordering appears to fully encompass the lamellar bilayer region of the phase diagram. There we find bilayers that possess distinct hexagonal closed packed ordering of nanoparticles as shown in Figure 2d, as compared to the liquid-like layers found in the PLH phase shown in Figure 2b.

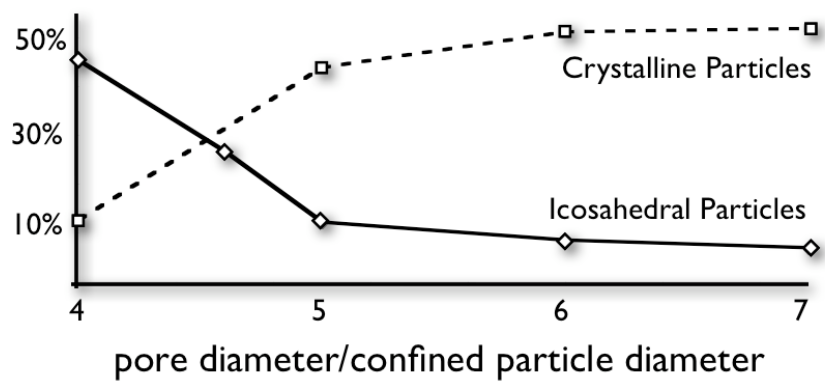

*Figure 4:* Percentage of icosahedral and crystalline nanoparticles under cylindrical confinement as a function of pore diameter. Icosahedral ordering is favored for diameters less than 5, which corresponds to the diameter of the tubes formed in the gyroid and cylinder phases.

*The role of confinement:*
It is known that small clusters of LJ particles will favor icosahedral packing; however as the system size approaches bulk behavior, such ordering is lost in favor of face-centered cubic and hexagonally close-packed crystals since icosahedra cannot tile Euclidean 3D space. The presence of icosahedral clusters in our systems is therefore surprising, since the domains formed contain large numbers of particles. The bulk phases that exhibit strong icosahedral ordering of nanoparticles, namely the H, DG, and PLH phases, have in common that the nanoparticle-rich domains are shaped like cylindrical tubes. For example, the nanoparticle rich domain in the H phase is comprised of cylinders, the DG phase has interconnected cylindrical tubes, and the perforations in the PLH phase induce curvature, creating sheets composed of interconnected tubular structures. We observe that there is little penetration of nanoparticles into the polymer-rich domain, thus the boundary between nanoparticle-rich and polymer-rich domains can be thought of as a confining surface. It appears that confining nanoparticles into tubular domains, as a result of microphase separation, allows for the formation of icosahedral clusters. To test this, we performed simulations in which we confined LJ particles of diameter $2\sigma$ within cylindrical pores of various diameters with $\phi = 0.25$ and $T^*=0.2$. The interaction between the particles and the walls of the pore were modeled by a Weeks-Chandler-Anderson potential. Examination of the dimensionless pore diameter, $d^*$ = pore diameter/confined particle size, shows a strong presence of icosahedral particles with both full and partial coordination at $d^*$ less than 5, and an associated decrease in crystalline particles (Figure 4). This model cylindrical pore system relates well to the tubular H and DG phases we observe where for H, $\phi \sim 0.2\text{-}0.3$ and $d^*\sim4.5\text{-}5$, and for DG, $\phi \sim 0.3$ and $d^*\sim4.5\text{-}5$.

Our results demonstrate the important role played by the local packing of nanoparticles within microphase-separated domains on the stability of the bulk structure for a tethered nanoparticle system. We observe a particularly interesting interplay between the local packing of nanoparticles and domain shape, which stabilizes the double gyroid phase and may potentially be exploited in other tethered nanoparticle systems to stabilize even more complex structures. More generally, we find that cylindrical confinement, whether from hard walls or as a result of microphase separation, can be used to promote icosahedral ordering between attractive spheres. Both findings have important implications for a variety of systems, including nanoparticle, colloidal, surfactant and biopolymer systems.

***Acknowledgements:*** We thank R.G. Larson, Z-L. Zhang, and N.A. Kotov for helpful discussions. Financial support for this work was provided by the U.S. Department of Energy, Grant No. DE-FG02-02ER46000 and the U.S. Department of Education GAANN Fellowship.